# Optical and Mid-Infrared Observations of the Planetary Nebula NGC 6781

J.P. Phillips[1], Gerardo Ramos-Larios[1], Martín A. Guerrero[2]

[1]Instituto de Astronomía y Meteorología, Av. Vallarta No. 2602, Col. Arcos Vallarta, C.P. 44130 Guadalajara, Jalisco, México   e-mails : jpp@astro.iam.udg.mx; gerardo@astro.iam.udg.mx

[2]Instituto de Astrofísica de Andalucía, CSIC. Glorieta de la Astronomía s/n, 18008 Granada, Spain
email: mar@iaa.es

**Abstract**

Although the planetary nebula (PN) NGC 6781 appears to possess an elliptical morphology, its kinematic and emission characteristics are in many ways unusual, and it is possible that it may represent a bipolar source oriented close to the line of sight. We shall present deep imaging of this nebula in [OIII] $\lambda5007$, H$\alpha$, and [NII] $\lambda6584$, and using broad-band (F555W and F814W) filters centred at $\lambda8269$ and $\lambda5252$. These were taken with the 2.56 m Nordic Optical Telescope (NOT) and Hubble Space Telescope (HST). This is combined with mid-infrared (MIR) imaging and spectroscopy acquired with the Spitzer Space Telescope (Spitzer), and near infrared spectroscopy deriving from the Infrared Space Observatory (ISO). These reveal details of the complex [NII] structure associated with extended shell emission, perhaps associated with highly inclined bipolar lobes. We also note the presence of narrow absorbing filaments and clumps projected against the surface of the envelope, components which may be responsible for much of the molecular emission. We point out that such clumps may be responsible for complex source structure in the MIR, and give rise to asymmetries in emission along the major axis of the source. Although most of the MIR $H_2$ v=0-0 emission is clearly concentrated in the bright interior shell, we shall also find evidence for extended emission to the north and south, and determine rotational excitation temperatures of order ~980 K.

Key Words: (ISM:) planetary nebulae: individual: NGC 6781 --- ISM: jets and outflows --- infrared: ISM --- ISM: lines and bands

# 1. Introduction

The planetary nebula (PN) NGC 6781 appears to possess an elliptical morphology at optical wavelengths, although it also has several emission and kinematic properties which are highly unusual for sources of this type. Levels of CO and $H_2$ emission appear to be extremely high compared to many other PNe, for instance (see e.g. Bachiller et al. 1993; Bell et al. 2007 for CO and $CO^+$, and Rosado & Arias 2003; Arias & Rosado 2002; Hiriart 2005 and Zuckerman et al. 1990 for $H_2$ emission), whilst the ionized and molecular shells have similar dimensions and structures. There is also evidence that that $H_2$ S(1) emission extends slightly outside of the [NII] shell (e.g. Zuckerman et al. 1990), whilst the expansion velocities of CO and $H_2$ (Bachiller et al. 1993; Hiriart 2005) are much larger than for the ionized regime (Weinberger 1989; Arias & Rosado 2002).

Mapping and spectroscopy of the envelope also turns up several surprises, and suggests that the interior ionized shell takes the form of a tilted cylindrical structure (Schwarz & Monteiro 2006); a situation which also applies for the $H_2$ envelope (Hiriart 2005), and is very similar to the truncated ellipsoidal modelling for CO J=2-1 (Bachiller et al. 1993). This, together with the evidence for $H_2$ emission, and southerly and northerly extensions to the primary emission shell (see e.g. Rosado & Arais 2003; Mavromatakis et al. 2001), may imply that the source is similar to NGC 7026, and represents a bipolar outflow oriented nearly pole-on to the line of sight (see e.g. Arias & Rosado 2002; Kastner et al. 1994, 1996).

We present deep narrow-band images of NGC 6781 obtained with the Nordic Optical Telescope (NOT), supplemented by archival observations acquired using the Hubble Space Telescope (HST). These will be used to investigate the optical structure and properties of the envelope, including evidence for exterior emission arcs and shell absorption. We also present archival imaging and spectra acquired with the Spitzer Space Telescope (SST) and Infrared Space Observatory (ISO), which will be used to investigate the excitation temperature of $H_2$, and nature and distribution of the mid-infrared (MIR) emission.

## 2. Observations

### 2.1 Optical Observations with the Nordic Optical Telescope

Narrowband images in the [NII], H$\alpha$ and [OIII] emission lines were obtained on 2009 June 22-23, using the ALFOSC (Andalucia Faint Object Spectrograph and Camera) mounted on the 2.56-m NOT; a facility which is based at the Observatorio de Roque de los Muchachos (ORM) in La Palma, Spain. The camera contained an 2048x2048 E2V CCD with pixel scale of 0.19 arcsec, and field of view (FoV) of 6.5x6.5 arcmin$^2$, whilst the filters had central wavelengths and bandwidths of $\lambda_C$ = 6589 Å, $\Delta\lambda$ = 9 Å (for [NII]), $\lambda_C$ = 6568 Å $\Delta\lambda$ = 8 Å (H$\alpha$), and $\lambda_C$ = 5007 Å $\Delta\lambda$ = 8 Å ([OIII]). The seeing was in all cases measured to be ~1.2 arcsec. Two frames with integration times of 450 seconds were taken for each of the filters, leading to total exposure periods of 900 s. The data were bias-subtracted and flat-fielded using twilight flats, and employing standard IRAF[1] routines.

### 2.2 Broad-Band Imaging with the HST

The HST observations[2] of NGC 6781 were obtained using the WFPC2 (Holtzman et al. 1995; see also the WFPC2 Instrument Handbook, Biretta 1996), took place on the 24/07/1995, and correspond to HST program 6119 (Snapshot Survey for Companions of Planetary nebula Nuclei; P.I. Howard Bond). The imaging results were downloaded using the HST archive search page at http://archive.stsci.edu/hst/search.php, from which we have chosen continuum exposures taken with filter F555W (mean wavelength $\lambda_m$ = 5252 Å, effective bandwidth $\Delta\lambda$ = 1222.5 Å), having a total exposure time of $\Delta t$ = 180 s; and filter F814W ($\lambda_m$ = 8269 Å, $\Delta\lambda$ = 1758.0 Å, $\Delta t$ = 350 s). These filters are selected to show various absorption features in the nebular shell.

---

[1] IRAF, the Image Reduction and Analysis Facility, is distributed by the National Optical Astronomy Observatory, which is operated by the Association of Universities for Research in Astronomy (AURA) under cooperative agreement with the National Science Foundation.

[2] Based on observations made with the NASA/ESA Hubble Space Telescope, and obtained from the Hubble Legacy Archive, which is a collaboration between the Space Telescope Science Institute (STScI/NASA), the Space Telescope European Coordinating Facility (ST-ECF/ESA) and the Canadian Astronomy Data Centre (CADC/NRC/CSA).

## 2.3   Spitzer Imaging in the MIR

The Spitzer[3] imaging of NGC 6781 with the Infrared Array Camera (IRAC; Fazio et al. 2004) was taken on 20/04/2004 as part of Program 68 (Studying Stellar Ejecta on the Large Scale using SIRTF-IRAC; P.I. Giovanni Fazio), and the results have been processed as described in the IRAC Instrument Handbook Version 1.0, February 2010 (available at http://ssc.spitzer.caltech.edu/irac/iracinstrumenthandbook/IRAC_Instrument_Handbook.pdf). The resulting post-Basic Calibrated Data (post-BCD) are relatively free from artefacts; well calibrated in units of MJy sr$^{-1}$; and have reasonably flat emission backgrounds. The observations employed filters having isophotal wavelengths (and bandwidths $\Delta\lambda$) of 3.550 $\mu$m ($\Delta\lambda$ = 0.75 $\mu$m), 4.493 $\mu$m ($\Delta\lambda$ = 1.9015 $\mu$m), 5.731 $\mu$m ($\Delta\lambda$ = 1.425 $\mu$m) and 7.872 $\mu$m ($\Delta\lambda$ = 2.905 $\mu$m). The normal spatial resolution for this instrument varies between ~1.7 and ~2 arcsec (Fazio et al. 2004).

## 2.4  Spitzer Spectroscopy in the MIR

Spectroscopy of NGC 6781 was obtained through Spitzer Program 1425 (SIRTF IRS Calibration Program; P.I. Lee Armus) on 3/2/2007, in which 7.4-14.2 $\mu$m observations were acquired using the Short-Low (SL) module of the Spitzer Infrared Spectrograph (IRS; Houck et al. 2004). The five positions observed (A to E) are indicated in Fig. 1, where one side of the slits corresponds to first order SL1 (7.4-14.5 $\mu$m) results (denoted with A1, B1 & etc.), and the other to second order SL2 (5.2-7.7 $\mu$m) spectroscopy (A2, B2 etc.). The respective aperture sizes are 3.7x57 arcsec$^2$ (for SL1), and 3.6x57 arcsec$^2$ (for SL2). It will be seen that the slits sample areas outside of the nebular shell (A1 & E2); emission immediately outside of the major axis limits of the nebular shell (A2, B1 to the south, D2 and E1 to the north), and central portions of the primary shell itself (B2, C1, C2, and D1).

## 2.5   ISO Spectroscopy in the NIR

---

[3] This work is based, in part, on observations made with the Spitzer Space Telescope, which is operated by the Jet Propulsion Laboratory, California Institute of Technology under a contract with NASA.

Finally, we make use of spectroscopy acquired by the Infrared Space Observatory (ISO)[4], a 60 cm f/15 telescope which was launched on 17/11/1995. This carried a Short Wave Spectrometer (SWS; de Grauuw et al. 1996) operating between 2.4 and 45 $\mu$m, and had aperture sizes varying from 14x20 arcsec$^2$ for $\lambda$ = 2.38$\rightarrow$12.1 $\mu$m, 14x27 arcsec$^2$ between $\lambda$ = 12$\rightarrow$29 $\mu$m, and 20x22 arcsec$^2$ between 29 and 45.3 $\mu$m. The present results were taken on 9/04/1997 as part of program 47901509, had an exposure period of 1192 s, and were centred on $\alpha$(2000.0) = 19h 18m 28.04s, $\delta$(2000.0) = +06° 32' 18.1".

## 3. Spitzer Spectroscopy of NGC 6781

There are no adequate background observations for the Spitzer spectroscopy of NGC 6781, since all of the slits contain nebular emission. However, levels of flux are low for positions A1 and E2, and we have subtracted spectra at these locations from the rest of the data. This permits us to remove systematic glitches and biases from the results, and study variations in emission along the major axis of the source (see Fig. 2, where we show the 1$^{st}$ and 2$^{nd}$ order results). Note however that whilst these variations are much clearer than in the raw data results, the fluxes for certain of the transitions are somewhat reduced. The peak flux of the v = 0-0 S(5) transition of H$_2$ is smaller by ~0.018 Jy, for instance, corresponding to ~7% of the line strength for position C2. Similarly, the strength of the S(4) transition appears more-or-less invariant with major axis offset, and the subtraction of spectrum A1 causes it to disappear from Fig. 2.

Apart from these caveats, all of the fluxes are likely to be reliable, and inspection of Fig. 2 reveals several interesting trends. The first is that there is a narrow emission band peaking at ~11.27 $\mu$m, likely corresponding to the C-H out-of-plane bending modes of neutral PAH molecules. This has a similar strength in all of the original spectra, implying that much of it may be associated with unrelated background emission. However, the "background" subtracted spectra in Fig. 2 show an increase in strength at positions C1 and D1 – a variation which suggests that the band is stronger at the centre, and likely to be intrinsic to the source. It is also interesting to note that there appears to be a stronger continuum close to positions C1 and D1, likely originating from Bremsstrahlung and/or warm dust continua.

---

[4] Based on observations with ISO, an ESA project with instruments funded by ESA Member States (especially the PI countries: France, Germany, the Netherlands and the United Kingdom) and with the participation of ISAS and NASA.

A similar trend applies to the low excitation [NeII] $\lambda$12.81 $\mu$m transition. This appears strong at positions C1 and D1, and is also reasonably intense at position E1. The line is however weak at B1, and almost non-existent at A1, suggesting that the transition is centrally concentrated upon the bright nebular shell, and drops-off quite steeply outside of this regime. A similar situation applies for [ArIII] $\lambda$8.99 $\mu$m and [SiIV] $\lambda$10.51 $\mu$m (the strongest of the IRS lines), although the absence of the latter transition at any but positions C & D suggests an even more concentrated level of emission.

The most dominant component, in terms of total levels of emission in Fig. 2, appears to correspond however to the v = 0-0 S(2)-S(7) transitions of molecular hydrogen. These appear strongest in the brighter portions of the nebular shell (B2-C1, C2-D1), and are weaker (but still strong) at distances < 60 arcsec north and south of the rim (A2-B1, D2-E1). The exterior A1 and E2 slit locations have very much less $H_2$ emission, although the S(4) and S(5) transitions appear to be detected in these regimes (see our comments above). There are also indications that S(2) emission may be present as well, although the peak appears to be kinematically shifted from that of the brighter central lines. Although such a difference in the lines merits further study, it is likely to derive from the reduced S/Ns of the fainter, outer transitions, and possible contamination by background emission; it would be unlikely to correspond to a real velocity shift.

This extension in $H_2$ emission is consistent with what has previously been observed in NIR S(1) mapping of the source, although it suggests that $H_2$ may be being detected to larger distances N and S of the elliptical shell, along the direction of the putative bipolar lobes mentioned in Sect. 1 (see also Sect. 6).

## 4. The Excitation Temperature of Molecular Hydrogen

It is well established that the $H_2$ molecule can be excited through a variety of mechanisms, including infrared fluorescence and collisional excitation. In the former case, absorption of a UV photon in the Lyman

& Werner bands leads to a rotational-vibrational cascade in the ground electron state (Black & van Dishoeck 1987). For this case, the UV photons may originate directly from the central star, or as a result of emission from strong shocks. On the other hand, where moderate velocity shocks lead to a warm post-shock cooling regime, then the $H_2$ transitions may arise through thermal excitation. Several analyses of PNe in the NIR suggest that both of these are important (see e.g. Hora et al. 1999; Hora & Latter 1994; Sahai et al. 1998; Ramos-Larios et al. 2008 for examples of thermal excitation, and Hora et al. 1999; Dinerstein et al. 1998; Hora & Latter 1996, and Luhman & Rieke 1996 for UV excitation).

It is often suggested that the ratio of the S(1) v = 1-0 and v= 2-1 transitions may represent a useful diagnostic for discriminating between these various mechanisms. S(1)2-1/S(1)1-0 is typically ~0.1-0.3 for shocks, and ~ 0.5 for fluorescent excitation. This has been used to suggest that the NIR v=1-0 transitions in NGC 6781 are shock excited, and used to determine total $H_2$ masses in the region of ~0.09-0.4 $M_\odot$ (Hiriart 2005; Arias & Rosado 2002; Rosado & Arias 2003; see also Zuckerman et al. 1990). However, it has also been noted that similarly low S(1)2-1/S(1)1-0 ratios can arise where UV excitation occurs in a dense nebular environment (Sternberg & Dalgarno 1989), so that this ratio is not the decisive parameter which has often been assumed in studies of this kind. Examples of high density UV excitation include M 2-9 and NGC 7027 (Hora & Latter 1994, Hora et al. 1999; Graham et al. 1993). Although [SII] and [ArIV] densities for NGC 6781 appear to imply smaller values of $n_e$ (~440-600 $cm^{-3}$; Mavromatakis et al. 2001; ~500 $cm^{-3}$; Stanghellini & Kaler 1989), these are weighted by the higher levels of emission arising from lower density gas – a component which represents the larger portion of the nebular mass. They may not be typical of densities in the $H_2$ excitation regime.

Alternatively, there are sometimes marked differences between the vibrational and rotational excitation temperatures, as well as the detection of $H_2$ transitions from higher vibrational levels. Both of these arise through overpopulation of the higher vibrational states, and represent indicators of probable UV fluorescence (Hora & Latter 1996; Shupe et al. 1998; Rudy et al. 2001). The present $H_2$ detections correspond to the v = 0-0 transitions alone, and this precludes us from evaluating such trends in the present source. On the other hand, the S/N of the individual and combined $H_2$ results is relatively high, and sufficient to permit accurate estimates of $T_{EX}$(rot).

For the present case therefore, where there is no change in the vibrational state v, the column density of a transition with rotational state J, relative to the v=0-0 S(2) transition having J=4, is given through

$$\frac{g_4 N(v,J)}{g_J N(0,4)} = \exp\left\{-\frac{E(v,J)-E(0,4)}{kT_{ex}}\right\} \quad\ldots\ldots\ldots\ldots(1)$$

where the left hand side of this equation is equivalent to

$$\frac{F(v',J')\nu_{0,2S(4)}A_{0,4\to 0,2}g_4}{F(0,4)\nu_{\Delta v,\Delta J}A_{v',J'\to v'',J''}g_J} \quad\ldots\ldots\ldots(2)$$

In this case, F(v′,J′) is the observed line flux, and $A_{v',J'\to v'',J''}$ is the transition probability. We have used this expression, together with Einstein A coefficients deriving from Turner et al. (1977) (see also the summary by Darren L. DePoy at http://www.astronomy.ohio-state.edu/~depoy/research/observing/molhyd.htm), to determine the population trends illustrated in Fig. 3.

It is evident that the data possesses a straightforward linear log-log trend, where the emission bars arise from averaging over differing sectors of the nebular shell – that is, they derive from variations in emission rather than from noise in the results. The reciprocal gradient of this trend implies an excitation temperature ~ 980 K. Such an estimate is similar to the values determined from previous analyses of PNe, which imply a range $T_{EX}$(rot) ≈ 450→2300 K (see e.g. Hora & Latter 1996; Hora et al. 1999).

## 5. MIR Mapping and Profiles for NGC 6781

It is apparent, from our consideration of the MIR spectra in Sect. 3, that emission in the 5.8 and 8.0 μm bands is affected by strong $H_2$ emission, the λ8.99 μm transition of [ArIII], and a possible underlying continuum which increases to longer MIR wavelengths. The situation at shorter

IRAC wavelengths is rather less clear, although given the weakness of the continuum in Fig. 2, it is probable that ionic and $H_2$ transitions are the primary contributors to the flux. Some evidence for this is present in the ISO spectrum in Fig. 4. Although this is the highest S/N segment of the NIR portion of the ISO observations, it is clear that the levels of noise leave a lot to be desired. It is nevertheless apparent that continuum levels are low, and emission is dominated by ionic transitions. The nature of the lines at ~ 2.755 and 2.938 $\mu$m is not entirely assured, and we base our identifications on the ISO lists at http://www.mpe-garching.mpg.de/iso/linelists/.

It is therefore clear that a broad range of transitions and continua may affect emission in the IRAC images described below, and complicate the interpretation of relative emission strengths.

Contour maps of NGC 6781 are illustrated in Fig. 5, in which the lowest contours are set at 3$\sigma$ background noise levels or greater. The intrinsic surface brightness $E_n$ is given by $E_n = A10^{(n-1)C} - B$ MJy sr$^{-1}$, where n is the contour level ($n$ = 1 corresponds to the lowest (i.e. the outermost) level), and B is background. Values for the parameters A, B & C are quoted in the caption to the figure. It would appear, from these maps, that the morphology is similar in all of the channels. There are however differences between the wavelengths which may arise as a result of the emission components discussed above (Sect. 3). In particular, the more extended, weak emission in the 5.8 and 8.0 $\mu$m channels may be associated with exterior components of $H_2$ emission. The latter is known to be very strong in the NIR and MIR, with some evidence for weak extension to the north and south (see e.g. Rosado & Arias 2003; Arias & Rosado 2002; Hiriart 2005; Zuckerman et al. 1990, and our discussion in Sect. 3). It is also interesting to note that the edge of the elliptical envelope is very sharply defined, and forms a higher surface brightness annulus. This feature has also been noted in various ionic transitions (see our discussion above), as well as at near infrared (e.g. $H_2$ S(1-0), see Sect. 1) and millimetric wavelengths (CO J = 2-1; e.g. Hiriart 2005; Bachiller et al. 1993). After applying Gaussian de-convolution using the point spread function, we find that the rim has a minor axis FWHM of 107.4 arcsec in [NII]; 105.6 arcsec in H$\alpha$; and 96.0 arcsec in [OIII]; values which compare with the equivalent 8.0 $\mu$m diameter of 114.2 arcsec. Very closely similar values are obtained for the raw (convolved) data sets as well. It is therefore clear that there is an evolution in the FWHM of the source, and that the 8.0 μm dimensions are larger than those for the optical regime. This would be consistent with appreciable $H_2$ emission in the MIR (see Sect. 3), and extension of this emission outside of the ionised regime (e.g. Zuckerman et al. 1990).

The major axis profiles through the source, illustrated in Fig. 6, were obtained after correcting for the effects of background emission; a component which is present in all of the bands, but appears to be particularly strong at 5.8 and 8.0 μm. This involved the removal of large background offsets, as well as much smaller linear gradients. We have also obtained profiles of the 8.0μm/4.5μm, 5.8μm/4.5μm and 3.6μm/4.5μm flux ratios. Since there is very little emission by polycyclic aromatic hydrocarbons (PAHs) within the 4.5 μm channel, such ratios tend to show the distribution of PAHs where these are dominant within a source. Dust continua may also be important in the longer wave (5.8 and 8.0 μm) channels, and ionic and $H_2$ transitions are present in all of the IRAC bands (see Sect. 3).

Some care must be taken in interpreting the latter results, since scattering within the IRAC camera can lead to errors in relative intensities. The flux corrections for extended source photometry lead to maximum changes of ~0.91 at 3.6 μm, 0.94 at 4.5 μm, 0.66-0.73 at 5.8 μm and 0.74 at 8.0 μm (IRAC Instrument Handbook), although the precise values of these corrections also depends on the distribution of surface brightness in the source. In the face of these uncertainties, we have chosen to leave the surface brightness and flux ratio profiles unchanged. The maximum correction factors for the 8.0μm/4.5μm and 5.8μm/4.5μm ratios are likely to be > 0.8, but less than unity, and ignoring this correction has little effect upon our interpretation of the results.

We find a broad similarity of the trends in the differing IRAC bands; a very sharp peak at relative position RP = 53.8 arcsec, at the southerly limits of the elliptical shell; and an increase in surface brightness with increasing MIR wavelength. The opposite (northerly) limits of the shell are by contrast very much less well defined.

The lower panel of Fig. 7 shows the variation in the 8.0μm/4.5μm, 5.8μm/4.5μm and 3.6μm/4.5μm surface brightness ratios, from which it is apparent the profiles are mostly uniform, with deviations attributable to central and field star contamination. None of the ratios varies significantly over the inner portions of the source.

Similar trends are also noted along the minor axis of the source, although the E-W limits of the shell are in this case both very well defined, and represented by sharply defined emission peaks (Fig. 7, upper panel). The flux ratio trends are again relatively flat (Fig. 7, lower panel), although with minima in 5.8μm/4.5μm and 8.0μm/4.5μm at the positions of the emission peaks (RP ~ -49.3 and 52.4 arcsec), and a possible increase in the ratios to larger distances from the central star.

It is therefore clear that the fluxes for this source increase from shorter to longer MIR wavelengths, and that profiles are structurally similar in all of the IRAC bands. The latter characteristic is consistent with a dominance by $H_2$ and ionic emission, and the similarity between the visual and $H_2$ mapping of the source. Where warm dust emission is important, and contributes to the increase in fluxes noted in Figs. 6 & 7, then it is likely that the grains are primarily located within the main emission shell. This would be consistent with longer wave observations by Hawkins & Zuckerman (1989), which indicated that cooler dust components are located within ≈ 50 arcsec of the central star. The similarity between the 5.8μm/4.5μm and 8.0μm/4.5μm profiles would also suggest a comparable distribution of dust and gas.

Finally, dips in 5.8μm/4.5μm and 8.0μm/4.5μm at the positions of the bright emission peaks presumably arise from an increase in fluxes in the 4.5 μm channel; a consequence perhaps of stronger Brα emission at the limits of the source. Similarly, a corresponding dip in the 8.0μm/5.8μm ratios (not shown here) may imply a decrease in levels of [ArIII] emission. We note that the steep rise in 5.8μm/8.0μm and 8.0μm/4.5μm ratios at the minor axis limits of the envelope may be explainable where $H_2$ emission is again strong at 5.8 and 8.0 μm; extends outside of the ionised shell (see our comments above); and where the 4.5 μm channel has higher proportions of ionic and bremsstrahlung emission.

## 6. Imaging of NGC 6781 in the Visual and MIR

The visual imaging of NGC 6781 acquired using the NOT is presented in the left-hand panel of Fig. 8, whilst a corresponding Spitzer image of the source is shown in the right hand panel. Both of the images have the same spatial scales, positioning and orientation, and both have been processed using unsharp masking techniques. In the latter procedure, a blurred or "unsharp" positive of the original image is combined with the negative, leading to a significant enhancement in the "sharpness" of the image (see e.g. Levi 1974).

Several aspects of the images are of particular interest. In the first place, we note that the optical results show evidence for ionisation stratification, similar to that previously noted by Mavromatakis et al. (2001). [OIII] and H$\alpha$ transitions are dominant in the interior of the source, whilst [NII] is strongest in the exterior regimes. This is particularly clear in the normalised profiles in Fig. 9, where it is evident that there are marked differences between the transitions, and between the minor and major axis trends. In all cases, the central intensity of [NII] is of order ~10-15 % peak emission strength, whilst the corresponding value for the higher excitation lines is of the order of ~50-60 %.

Such trends might be explicable where the [NII] component arises in the outer regions of a cylindrical outflow structure (Schwarz & Monteiro 2006; Hiriart 2005), whilst the interior of the cylinder contains higher excitation gas. Similarly, the asymmetrical profiles along the major axis of the source are likely to be associated with the web of extinction structures noted in Fig. 8 – it appears that extinction is significantly larger to the north than it is towards the south. This result is consistent with the H$\alpha$/H$\beta$ mapping of Mavromatakis et al. (2001), who show that the ratio increases from 5.6 ± 0.2 in the south to 7.0 ± 0.5 in the north – implying a difference in mean extinction of $\Delta c \approx 0.26$ (or $\Delta A_V \approx 0.56$ mag). Interestingly enough, a similar disparity in extinction ($\Delta A_V \approx 0.37$ mag) appears capable of explaining the asymmetry of the [NII] profile – assuming, that is, that the intrinsic profile is symmetric. It would not however explain the MIR asymmetries in Fig. 6 – the levels of extinction, for this case, would be very much too small. It is possible however that individual condensations have much higher optical depths, and that the reddening is associated with low and high extinction regimes.

Finally, we note that such differences in $A_V$, taken in conjunction with the modelling of Hiriart (2005), Bachiller et al. (1993) and Schwarz & Monteiro (2006), suggest that most of the dust is located towards the outer surface of the cylinder. It is likely that we are looking through this dust towards the north, and at the internal surface of the cylinder towards the south.

A second aspect of the optical image, seen here more clearly that in previous observations of the source, is the presence of arc-like structures to the north and south, primarily visible in the [NII] transition. These protrude to distances ~40 arcsec outside of the central ellipse, and are similar to what might be expected for bipolar structures oriented at small angles to the line-of-sight. Such an interpretation is also consistent with the previous studies of the source (see Sect. 1; although note also the differing interpretation of Hiriart (2005)). We note however that the southerly arcs appear to emerge from behind the bright interior shell. This is the reverse of what might be expected where the southerly opening of the cylinder is oriented towards the observer, as suggested by the spatio-kinematic results of Bachiller et al. (1993) and Hiriart (2005), and the imaging and density mapping of Mavromatakis et al. (2001) and Schwarz & Monteiro (2006). This aspect of the source evidently requires further study (we are grateful to the referee, Prof. Joel Kastner, for pointing out this anomaly).

The Spitzer image of NGC 6781 is globally rather similar to that in the visible, with evidence for a bright elliptical rim with northerly and southerly envelope extensions. It is clear however that the similarity (for the most part) ends there. The Spitzer image shows that much of the emission derives from a network of filaments, and that the colour of the source is relatively uniform. Although the bright elliptical rim is somewhat whiter in appearance, testifying to stronger fractional components of lower wavelength emission, and there is some evidence for slighter redder emission (i.e. increased 5.8 μm fluxes) immediately outside of the rim, there is little evidence for the large disparities in colour noted in the NOT results.

This contrast between the NOT and Spitzer imaging is rather intriguing, and it is pertinent to ask what mechanisms might be responsible for the differences in structure.

Some answer to this question is provided in Fig. 10 (left-hand panels), where we show a further image of the NOT results processed using layer masking techniques (e.g. Montizambert 2002), designed to show even finer details of the structure than was possible in Fig. 9. We also illustrate broad-band imaging taken with the HST. The dark absorbing filaments are more obvious in both of the images, and seen to extend in ordered curves, lines and fragments over the entire area of the shell. The highest levels of absorption occur in the north-western portions of the envelope. It would also appear that the absorbing components have similar structures to those of the emission filaments in the MIR – it is as though we are detecting IR emission from the circum-nebular dust. This is not, necessarily, an implausible hypothesis, although careful registering of the images shows that it is not in fact the case. It is clear that the dark absorption filaments are bordered by [NII] emission sheaths, presumably arising due to ionisation stratification close to the absorbing cores. These [NII] sheaths also coincide with the Spitzer filamentary structures. Although the coincidence of the [NII]/Spitzer filaments is most clearly seen where one blinks spatially registered images of the source, we have also attempted to illustrate this trend in the right-hand panel of Fig. 10, where we show blow-ups of a bright MIR filamentary structure/visual absorption lane located ~23 arcsec west of the central star. The region of the Spitzer filament is outlined using an irregular white contour. It is apparent that the MIR emission is biased away from the absorbing lane, and towards the easterly fringe of [NII] emission. It is therefore possible that the $H_2$ and warm dust emission noted in the spectra of NGC 6781 (Sect. 3) may be arising close the HI/HII interfaces of the neutral (and dusty) mass components.

This, if it is the case, would also permit us to understand certain other aspects of the source. It is clear for instance that the $H_2$ S(1) and CO source structures are very similar to those in the visible, as noted previously in Sect. 1. This would make sense where there was sufficient extinction to prevent molecular dissociation (see e.g. Hasegawa 2003), and the molecular material was interlaced with the ionised gas, or located just outside of the HI/HII interface. That the latter is occurring in the present case is attested by the relative visual/MIR dimensions of the shell (see our comments in Sect. 5); the asymmetry in shell extinction noted above; and the fact that $H_2$ S(1) emission occurs just outside of the [NII] envelope (Zuckerman et al. 1990). The irregular nature of the extinction may also go some way to explaining the clumpy distribution of CO gas (Bachiller et al. 1993).

It is therefore clear that our present optical imaging of NGC 6781, when compared to Spitzer results, enables us to better understand the possible origins of MIR and $H_2$ emission.

## 7. Conclusions

NGC 6781 is a fascinating example of an elliptical PN for which the kinematic and emission properties appear to be anomalous. Levels of molecular emission are unusually high, whilst the interior shell appears to take the form of a tilted cylindrical structure. These and other characteristics suggest that the source is an inclined bipolar outflow.

We have presented deep NOT imaging of the nebula in [OIII] $\lambda$5007, H$\alpha$ and [NII] $\lambda$6584, and broad-band archival HST images at $\lambda$8269 and $\lambda$5252. We have also analysed archival Spitzer IRAC imaging in the 3.6, 4.5, 5.8 and 8.0 $\mu$m bands; IRS spectra in the range 5.2-14.5 $\mu$m; and an ISO SWS spectrum extending between 2.4 and 4.1 $\mu$m. The results show that [NII] emission extends ~40 arcsec north and south of the central shell, and possesses a complex multi-arc structure. By contrast, the interior portions of the shell are dominated by high excitation transitions. The morphology of the [NII] emission suggests that most of it arises in the walls of the cylindrical structure. We have also noted that there is evidence for complex absorption structures over the face of the nebula, likely associated with neutral filaments and condensations, and responsible for asymmetries in the emission structure. When one compares the variation in extinction with models of the nebular outflow, it seems probable that most of the dust is located in exterior portions of the cylindrical structure. It is also suggested that these high extinction condensations may be responsible for $H_2$ and CO emission; explain the similarity of the molecular and ionised structures; and be responsible for the "clumpy" appearance of the CO emission.

The filaments are also associated with enhanced [NII] emission, and it is suggested that these low excitation regimes are responsible for much of the MIR emission, accounting for the complex shell structures noted in the Spitzer observations.

IRS spectra of NGC 6781 show that $H_2$ v = 0-0 emission extends north and south of the principal envelope, and into the region occupied by the putative bipolar lobes. An analysis of mean trends in the line ratios shows rotational excitation temperatures to be of order $T_{EX}$ ~ 980 K. It remains unclear whether the emission arises due to shocks or UV excited fluorescence, and the situation may only become clearer through observations of further vibrational states.

**Acknowledgements**

GRL acknowledges support from PROMEP (Mexico), whilst MAG is partially funded by grant AYA2008-01934 of the Spanish Ministerio de Ciencia e Innovación (MICINN).

# Figure Captions

**Figure 1**

The location of five Spitzer IRS slit positions (labelled A-E) upon a 4.5 $\mu$m Spitzer image of NGC 6781. Although all of the slits are centred on the diagonal black line, and are therefore spatially overlapping, they have been displaced to the left and right in order to avoid confusion. Note how the slits have two separate portions (labelled 1 & 2) corresponding to the first and second orders of the short-low module.

**Figure 2**

"Background" subtracted spectroscopy for NGC 6781, where the results for positions A1 and E2 (Fig. 1) have been removed from more centrally located spectra. The vertical dashed lines represent the limits of the 5.8 and 8.0 $\mu$m IRAC channels, whilst the vertical dotted line indicates the limits the first and second order spectra (SL1 and SL2). The spectra are separated vertically by 0.15 Jy in order to avoid overlap and confusion. Similarly, we have associated pairs of first and second order results having similar spatial positions. The second order results at position C2 are roughly comparable to the first order results at position D1, for instance.

**Figure 3**

The mean variation in $H_2$ vibrational-rotational populations for the v = 0-0 S(2)-S(7) transitions. The results are based on averages of line strengths for all of the spectral positions in Fig. 1, whilst the "error bars" are related to variations between the spectra, rather than errors in the results. The least-square fit to the results implies rotational excitation temperatures of ~ 980 K.

**Figure 4**

ISO SWS results for the 2.4-4.1 $\mu$m spectrum of the central portions of NGC 6781. Although levels of S/N are modest compared to the Spitzer results, it appears that ionic transitions dominate the 3.6 $\mu$m IRAC band (represented by the red profile), and are likely to dominate the 4.5 $\mu$m channel as well (indicated by the purple profile).

**Figure 5**

Contour mapping of NGC 6781 in the four IRAC bands, where contour parameters [A, B, C] are given by [0.65, 0.1072, 0.344] at 3.6 $\mu$m, [0.4, 0.1338, 0.309] at 4.5 $\mu$m, [2.55, 0.0855, 2.082] for 5.8 $\mu$m, and [8.85, 0.0501, 8.433] at 8.0 $\mu$m. Note the presence of extended emission halos in the 5.8 and 8.0 $\mu$m channels. Although these appear much less obvious at 3.6 and 4.5 $\mu$m, this may be because of lower levels of surface brightness towards shorter wavelengths.

**Figure 6**

MIR profiles across the major axis of NGC 6781, where we show individual 3.6, 4.5, 5.8 and 8.0 $\mu$m trends (upper panel), and the 3.6$\mu$m/4.5$\mu$m, 5.8$\mu$m/4.5$\mu$m and 8.0$\mu$m/4.5$\mu$m ratios (lower panel). The direction and widths of the profiles are indicated in the inserted figures, where north is to the top (corresponding to the left-hand side of the emission profiles). Note how the individual MIR profiles are very closely similar, and lead to relatively flat variations in the flux ratios; deviations in these ratios are largely attributable to field and central star components. It will also be noted that the inner elliptical shell terminates fairly abruptly towards the southern limits of NGC 6781, leading to a sharp peaking in the levels of emission. The northerly shell limits are however much less well defined.

**Figure 7**

As for Fig. 6, but for the minor axis of NGC 6781. East is to the left-hand side of the inserted images, and right-hand side of the emission profiles. The minor axis limits of the shell are defined by two narrow emission peaks, and there is also evidence for a small decrease in 5.8$\mu$m/8.0$\mu$m and 8.0$\mu$m/4.5$\mu$m at the position of the peaks. We finally note evidence for an increase in these flux ratios outside of the primary shell, perhaps associated with $H_2$ emission at the HI/HII interface.

**Figure 8**

Optical and MIR imaging for NGC 6781. The left-hand panel shows the combined [OIII] (blue), H$\alpha$ (green) and [NII] (red) results taken with the NOT, whilst the right-hand panel shows combined 3.6 $\mu$m (blue), 4.5 $\mu$m (green) 5.8 $\mu$m (red) imaging deriving from Spitzer. The images are in register, and have the same orientation and spatial scale.

**Figure 9**

Normalised H$\alpha$ (dashed curves), [NII] (solid curves), and [OIII] (dot-dashed curves) NOT emission profiles along the major (upper panel) and minor (lower panel) axes of NGC 6781. The slit placements and widths are identical to those of the Spitzer profiles in Fig. 6, and are indicated in the inserted [NII] images. In the case of the upper profile, north is to the top of the inserted image, and to the left-hand side of the horizontal axis, whilst for the lower profile, east is to the left of the inserted image, and the right-hand side of the profile.

**Figure 10**

HST and NOT imaging of NGC 6781 (left hand panels), where the NOT image is similar to that illustrated in Fig. 9, but processed using layer masking, and the HST image is a combination of the 5252 Å (blue, green) and 8269 Å (red) broad-band filters (i.e. filters F555W and F813W). The two images are in register (i.e. have the same spatial scale, positioning and orientation), and show evidence for complex absorption structures. Many of the dark filaments and condensations are also bordered by [NII] emission. The right-hand panels show blow-ups of a region to the west of the central star, identified by the box within the upper left-hand panel. The white contour marks the limits of a Spitzer filamentary structure, which is seen to be biased towards the [NII] emission fringe bordering a dark absorption lane.

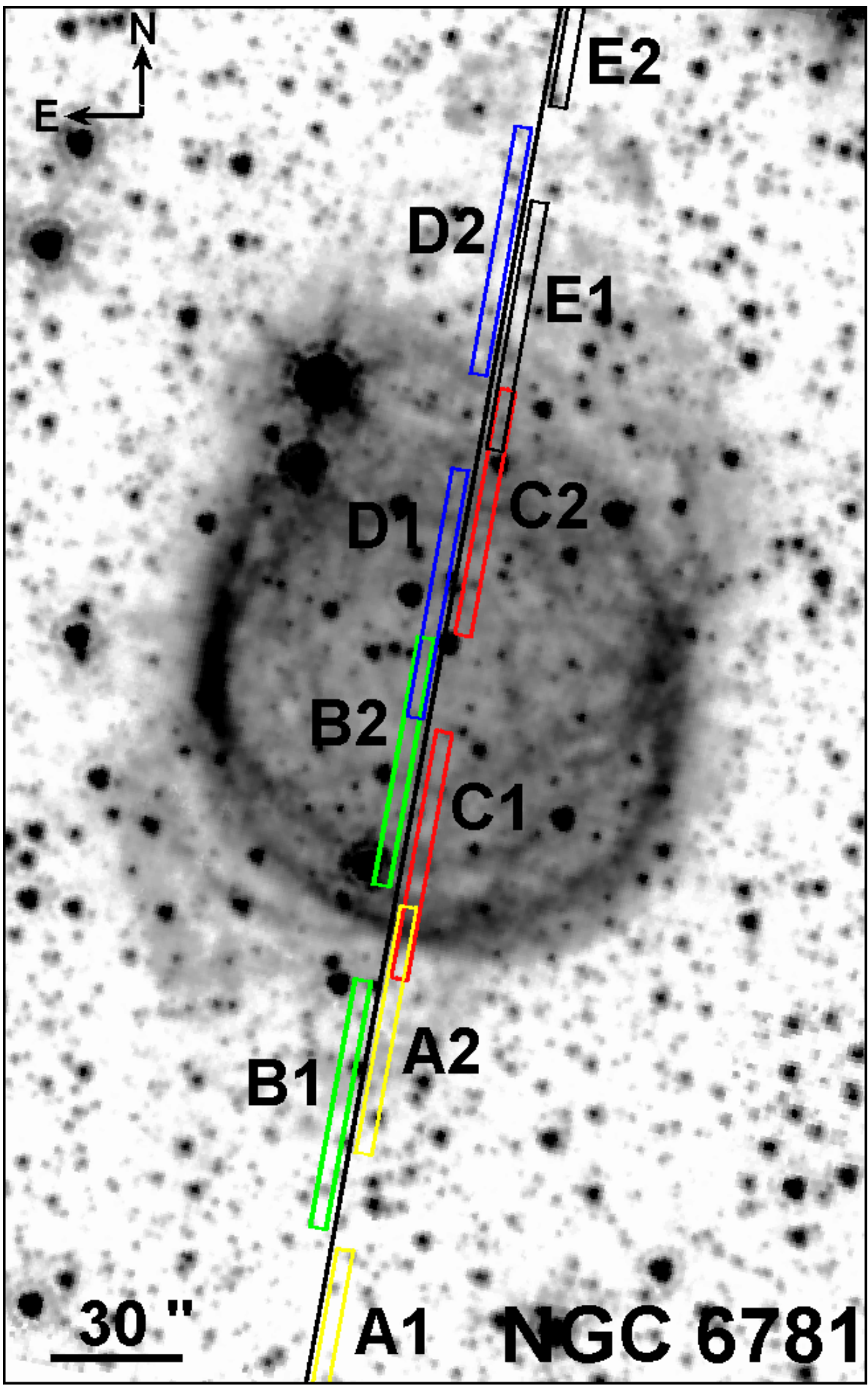

FIGURE 1

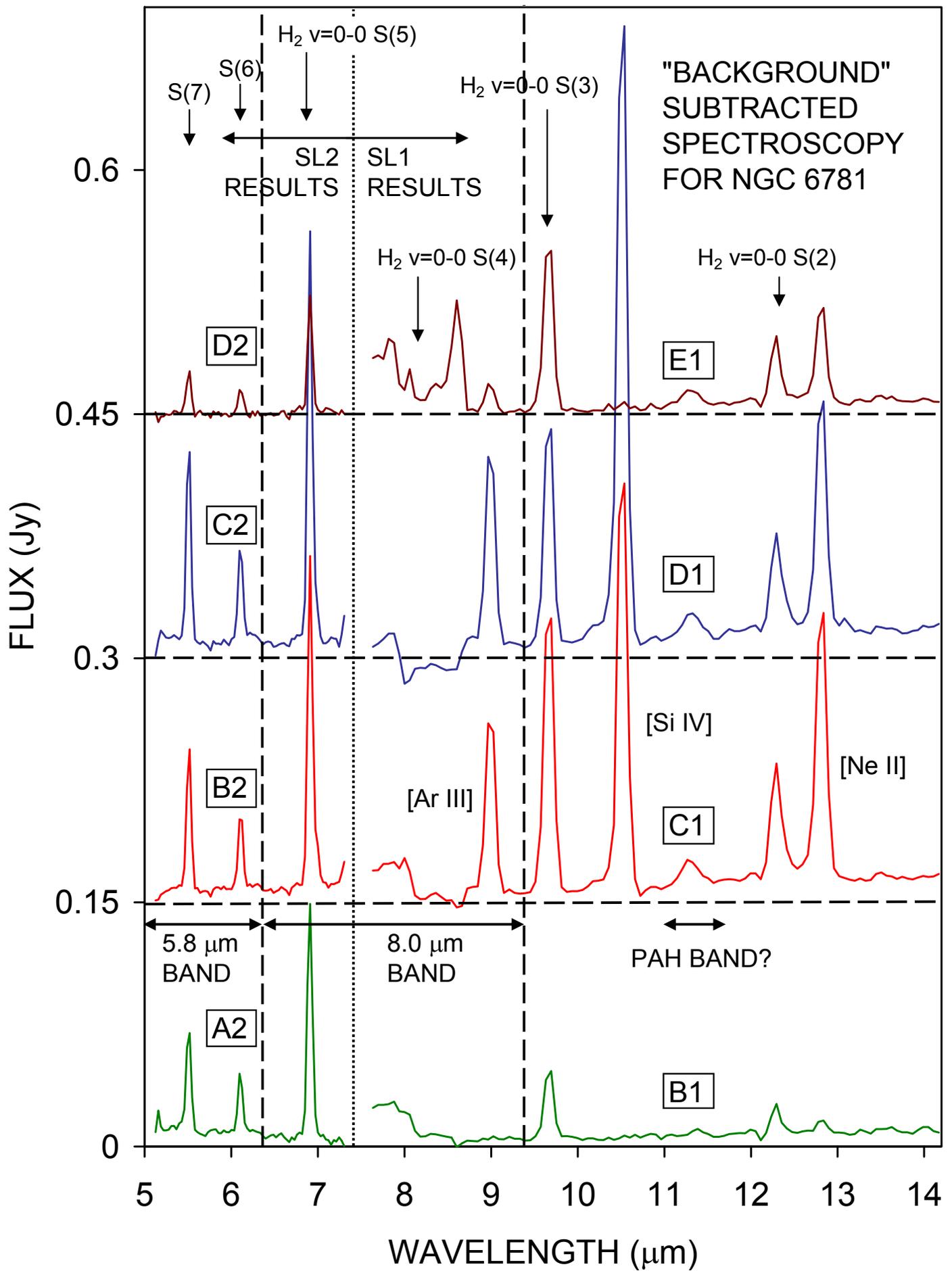

FIGURE 2

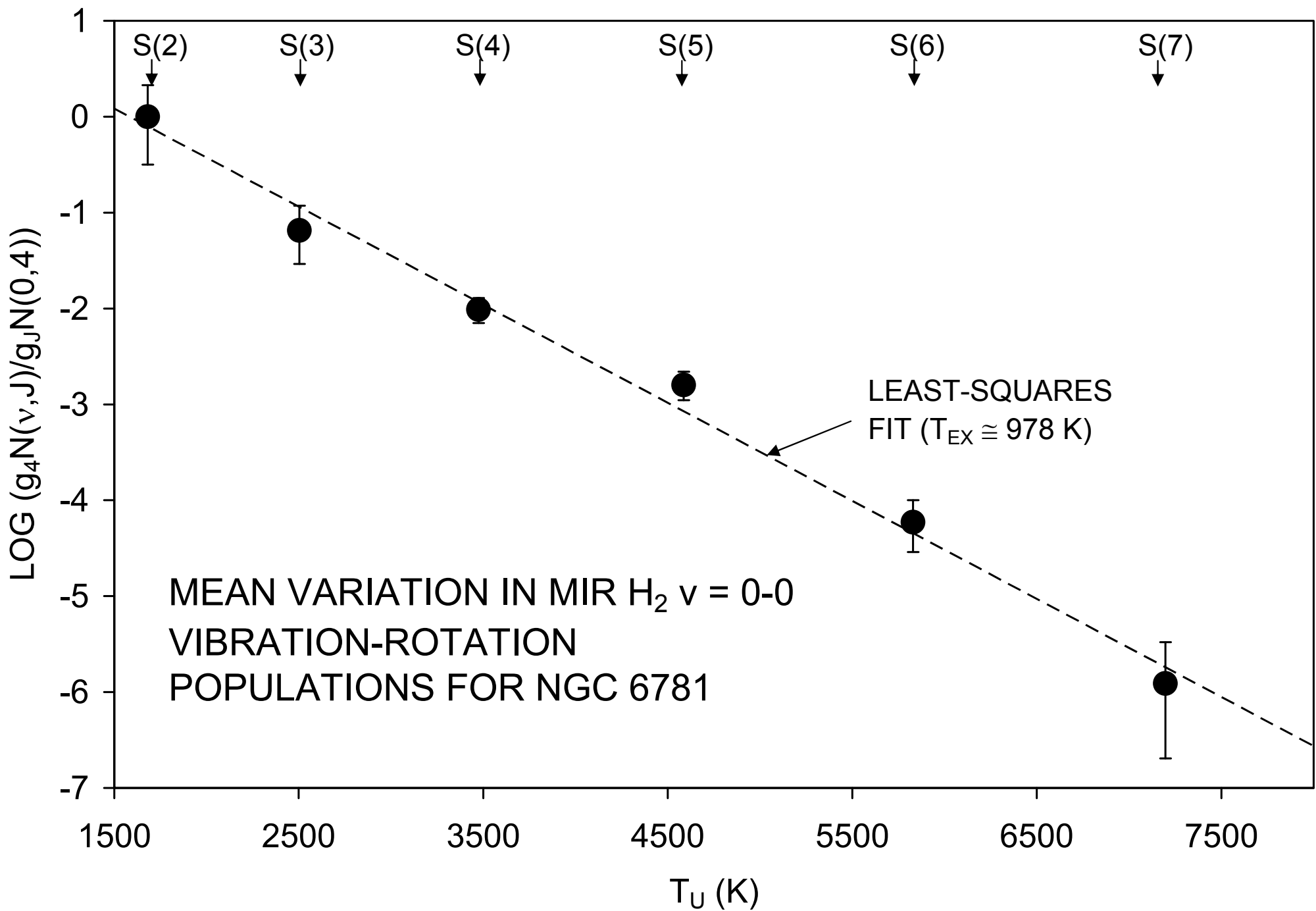

FIGURE 3

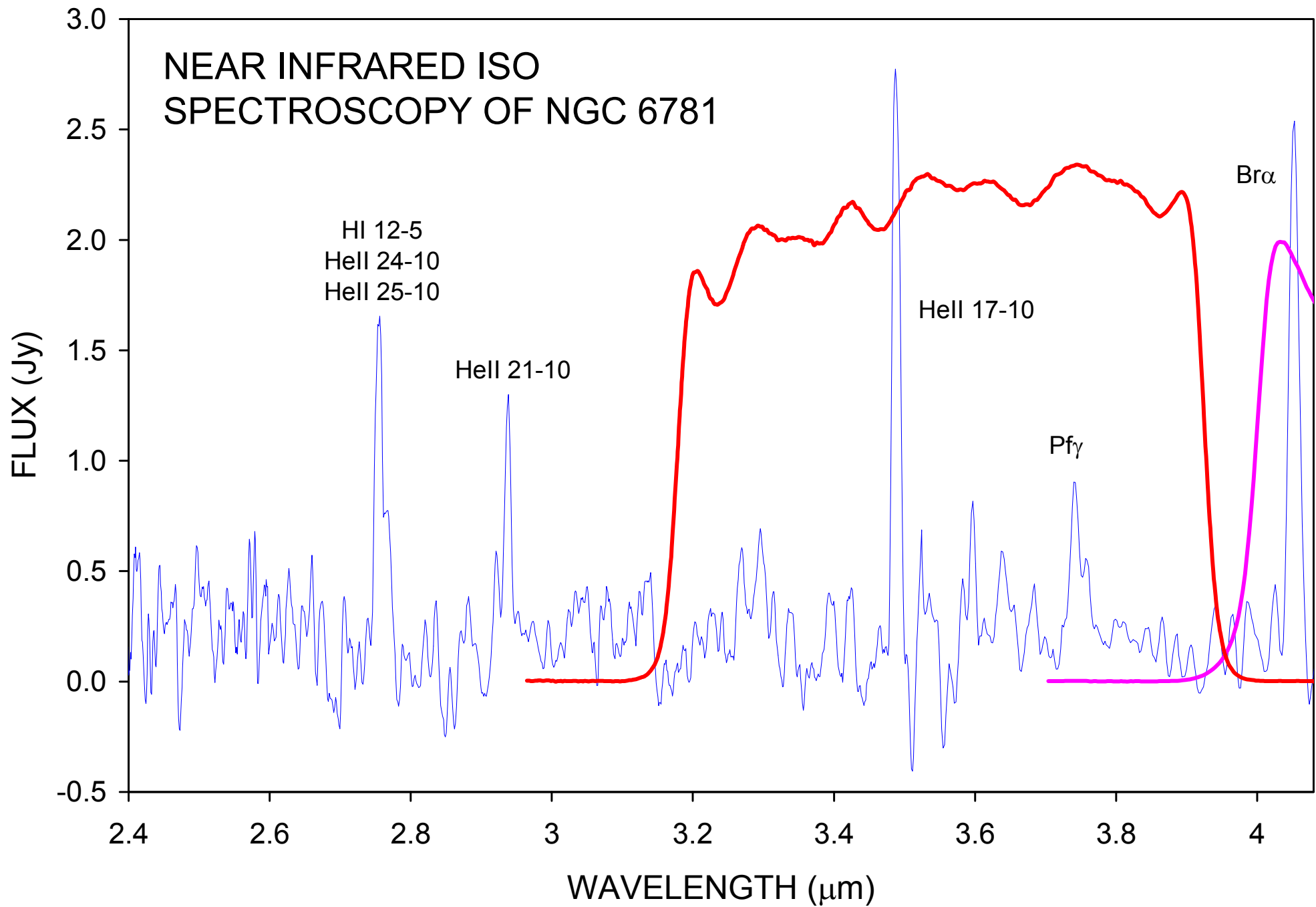

FIGURE 4

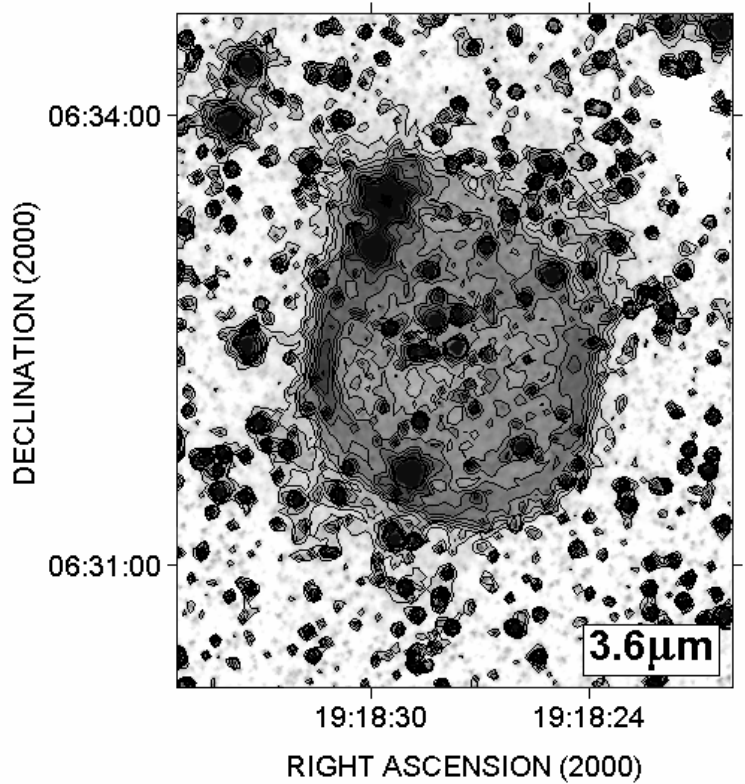
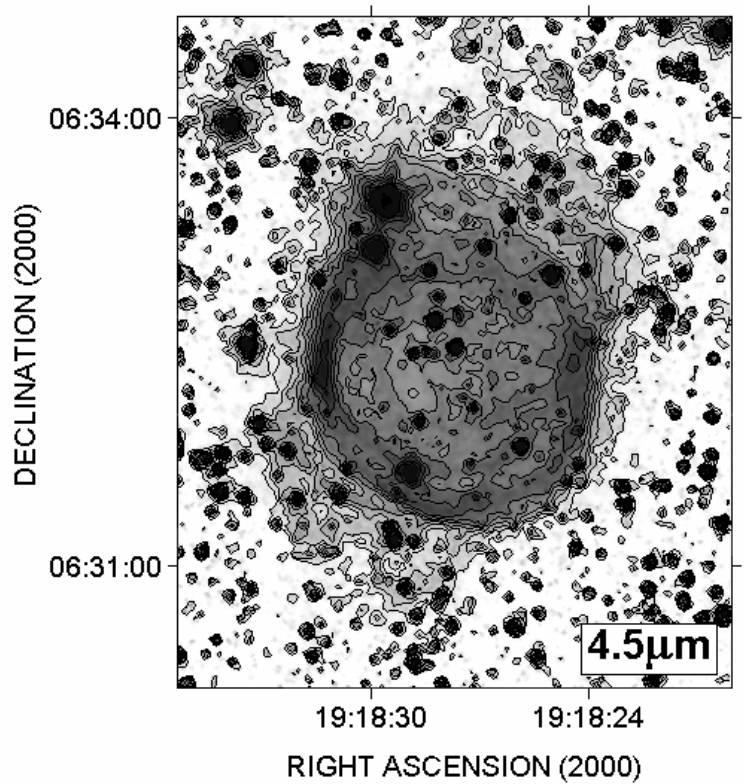
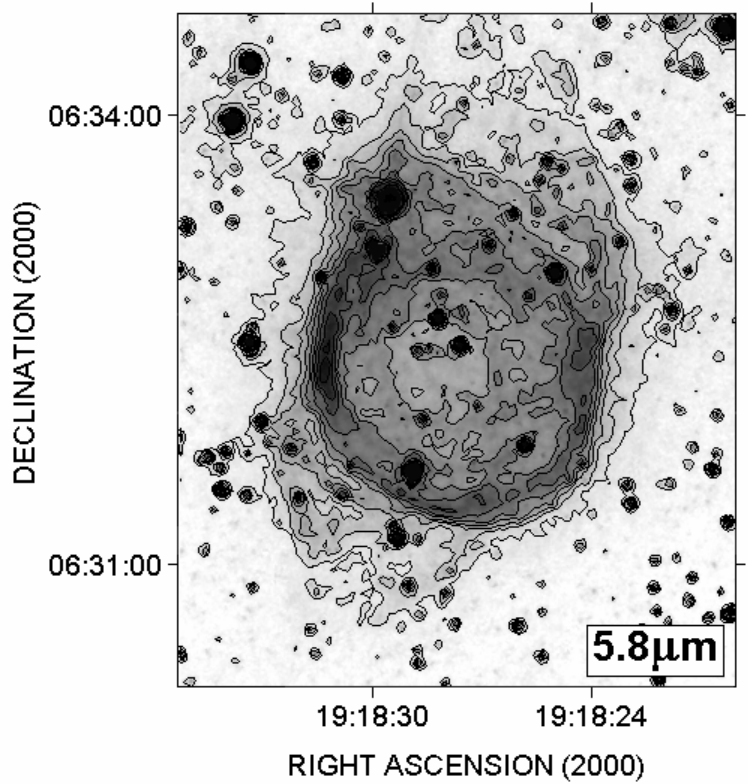
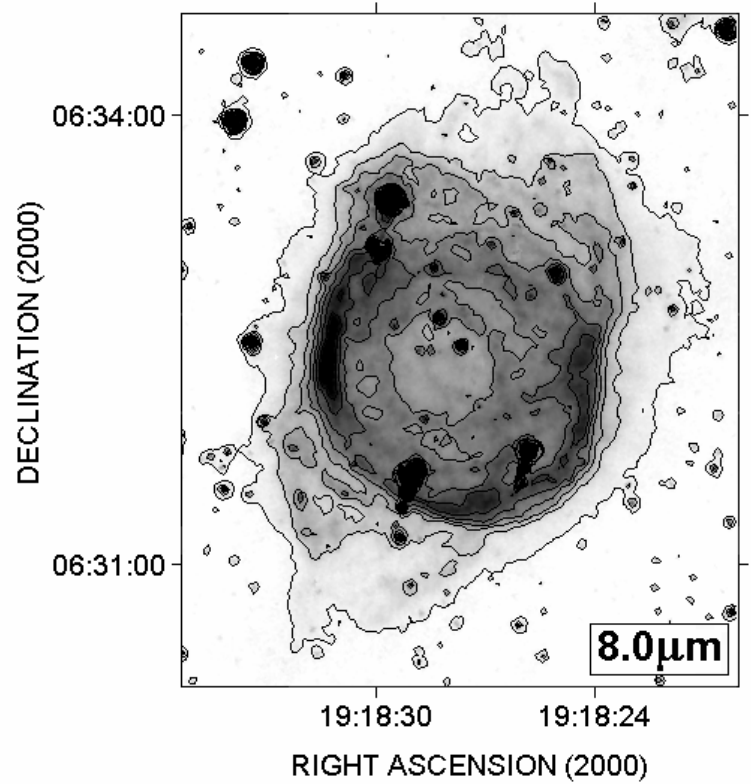

FIGURE 5

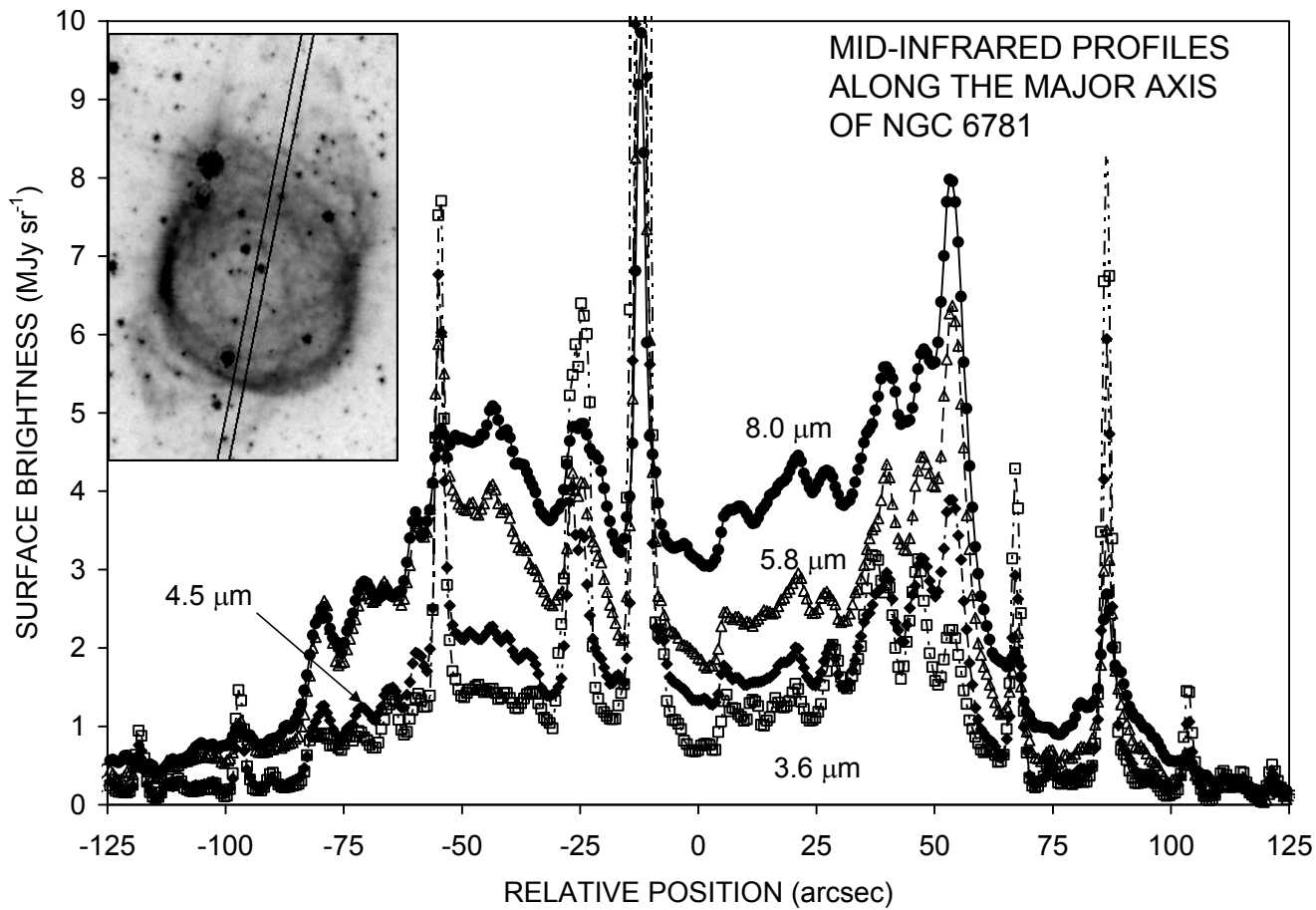

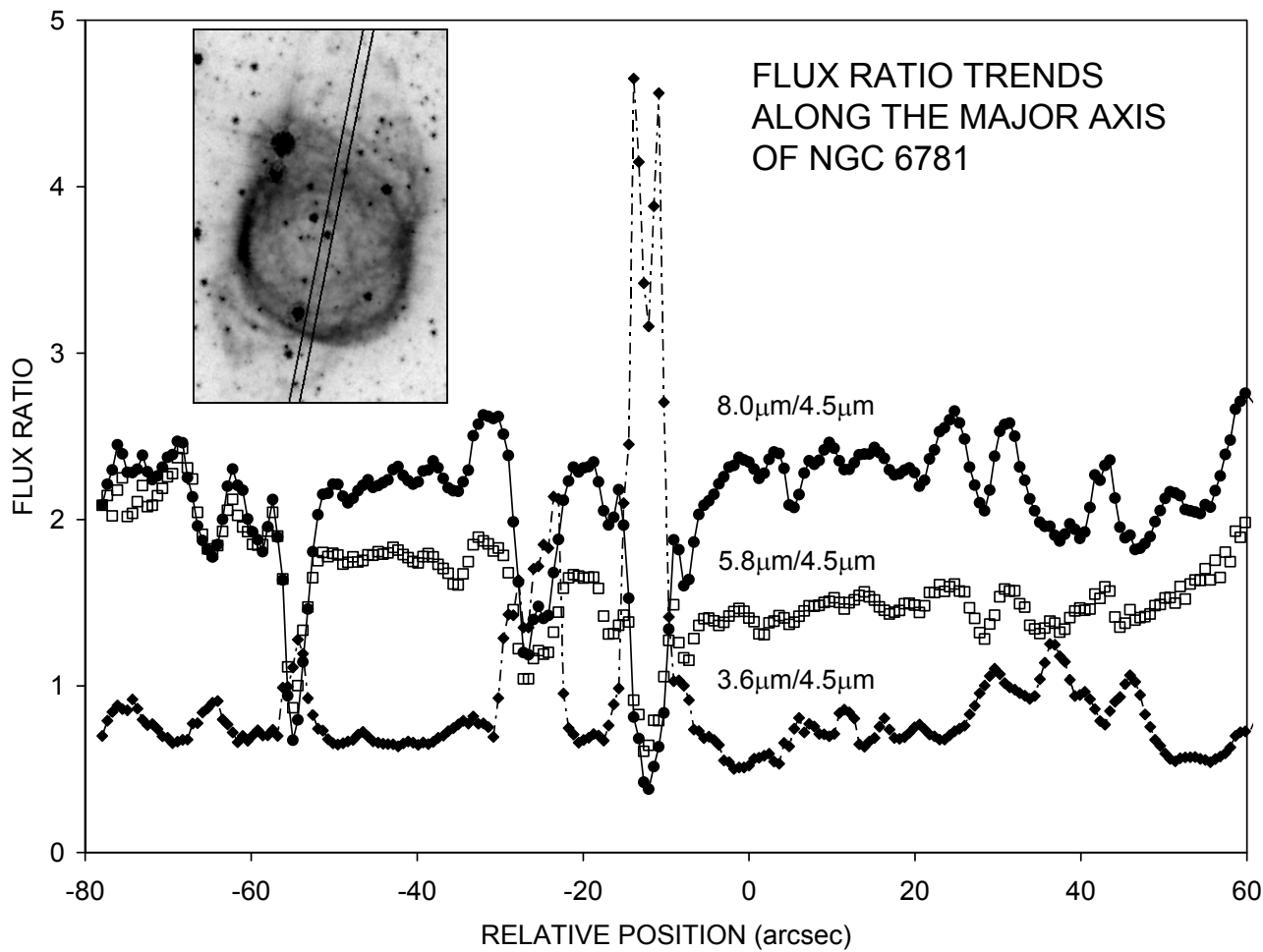

FIGURE 6

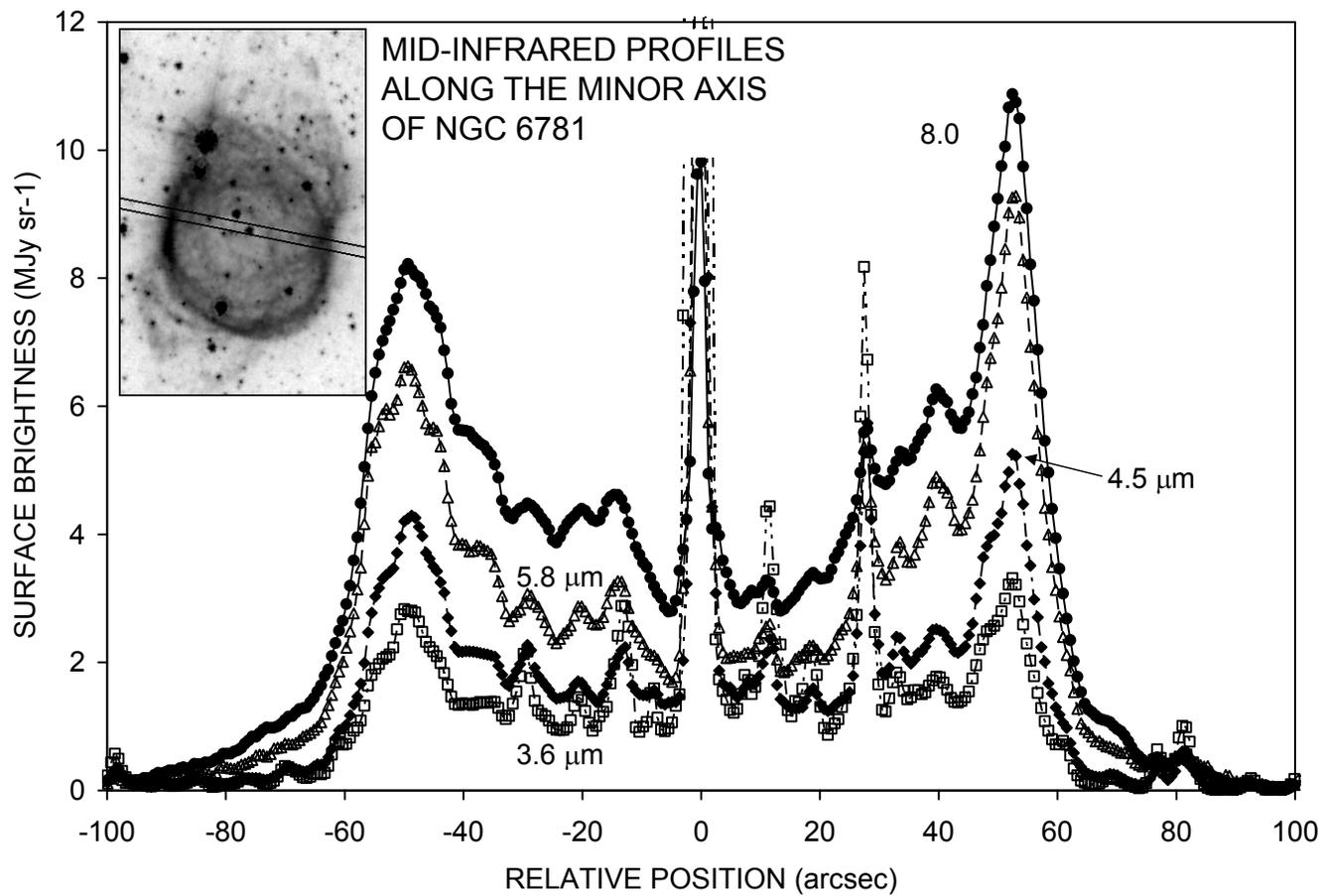

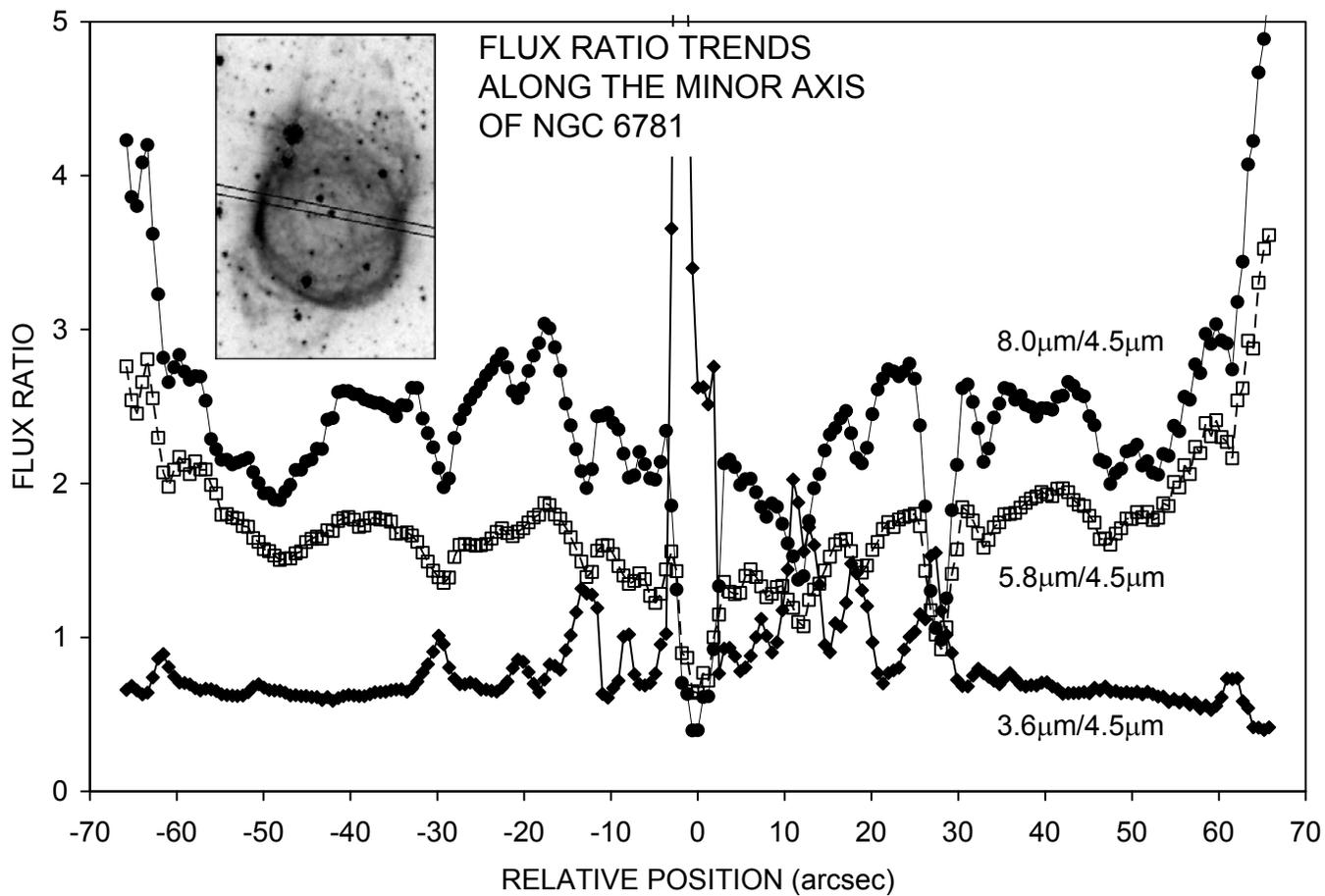

FIGURE 7

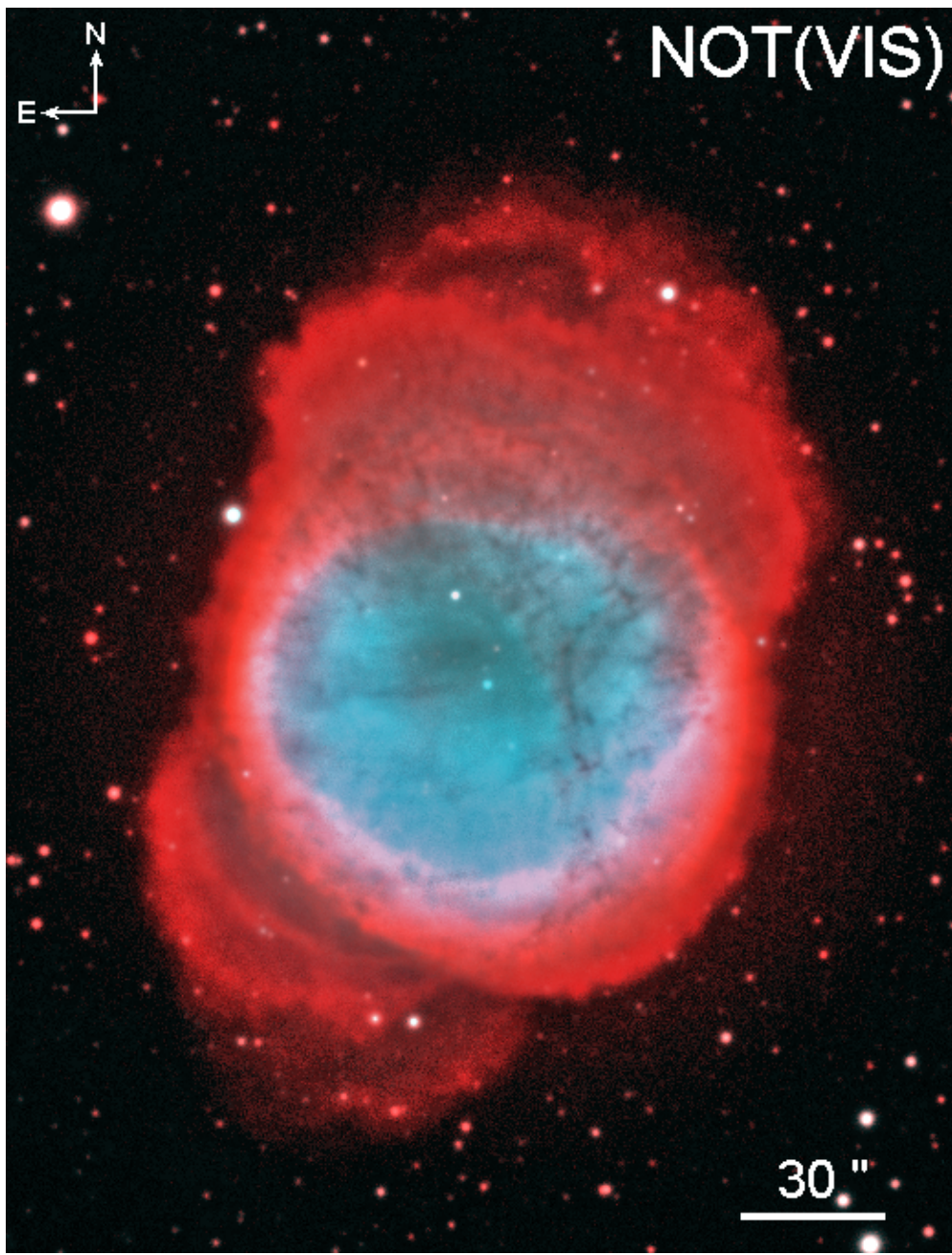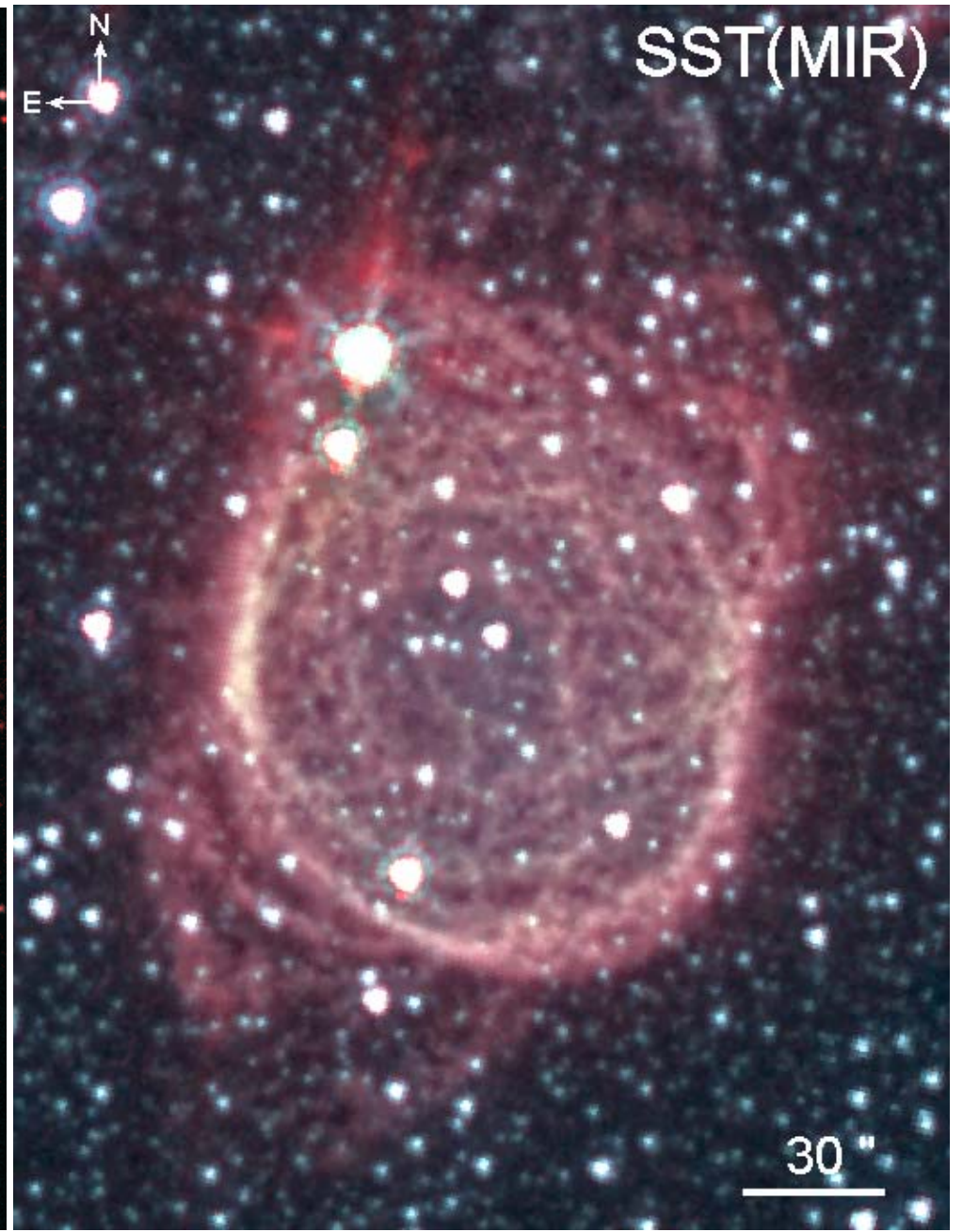

FIGURE 8

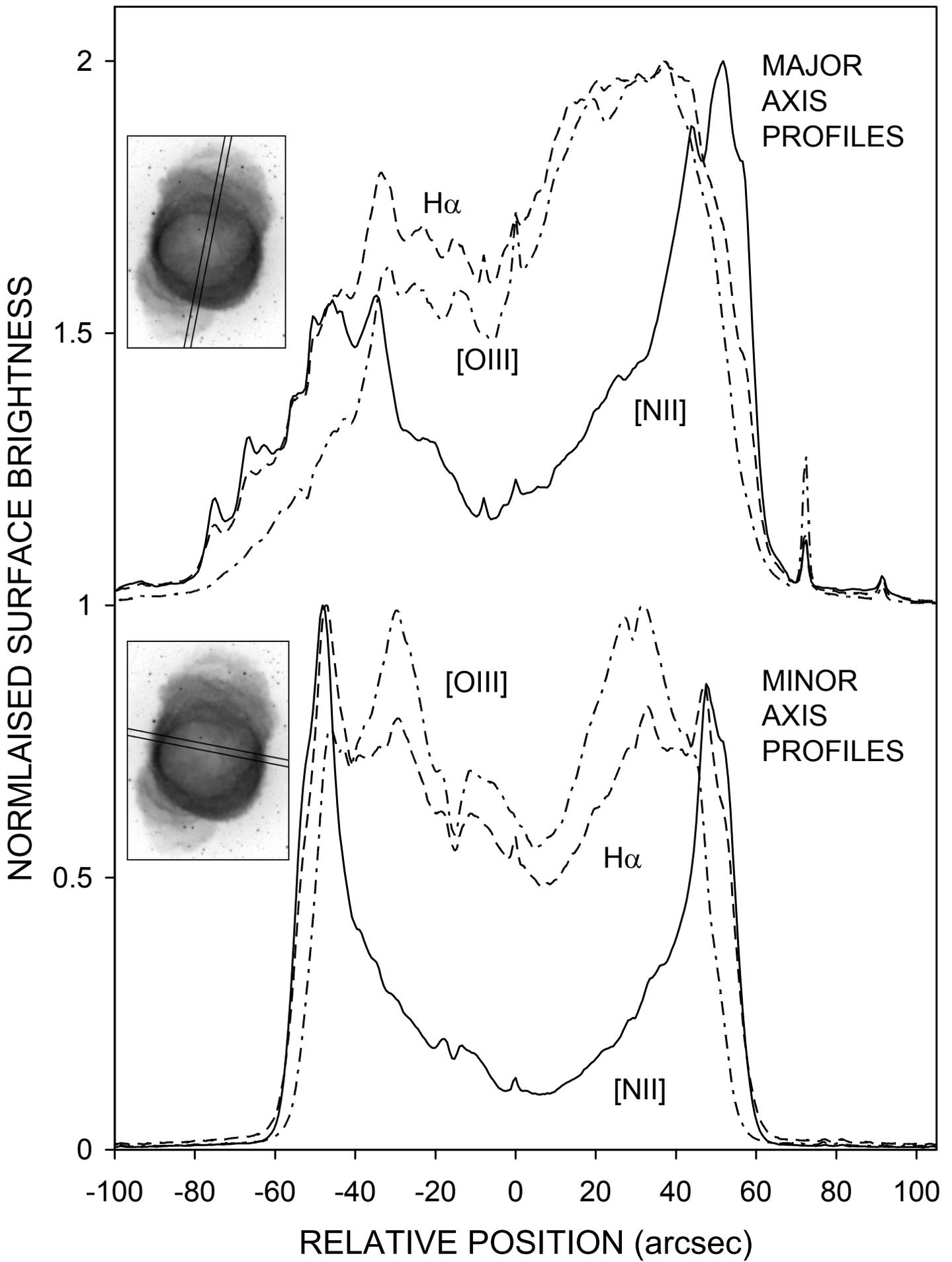

FIGURE 9

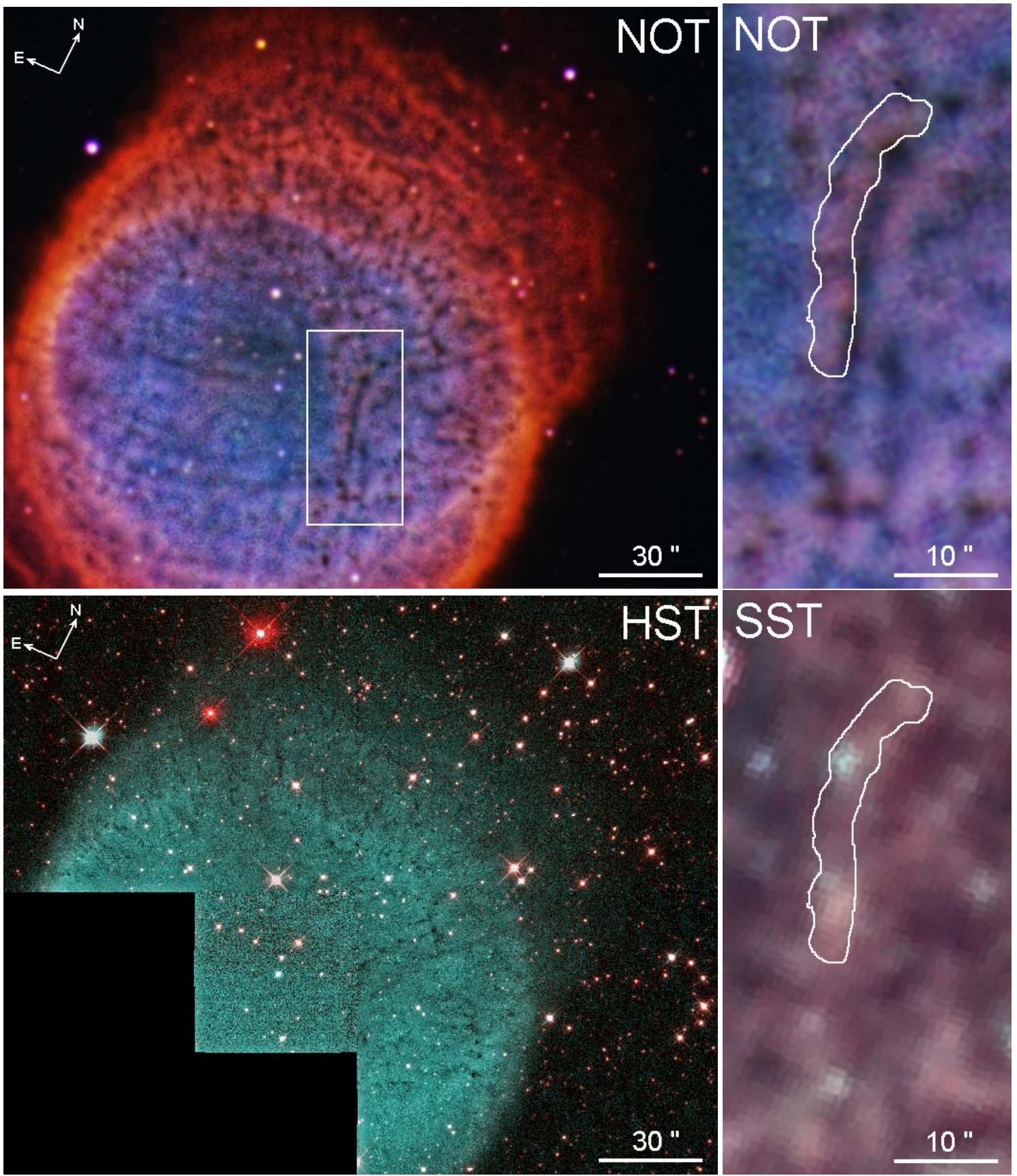

FIGURE 10

32